\documentclass{emulateapj}
\usepackage{graphicx, color}
\usepackage{longtable}
\usepackage{amssymb}
\usepackage{tabularx, blindtext}

\begin{document}
\shorttitle{SL analysis of El Gordo}
\shortauthors{Zitrin et al.}

\slugcomment{Submitted to the Astrophysical Journal Letters}

\title{A highly elongated prominent lens at $\MakeLowercase{z}=0.87$: first strong lensing analysis of El Gordo}


\author{Adi Zitrin\altaffilmark{1,*}, Felipe Menanteau\altaffilmark{2}, John P. Hughes\altaffilmark{2}, Dan Coe\altaffilmark{3}, L. Felipe Barrientos\altaffilmark{4}, Leopoldo Infante\altaffilmark{4}, Rachel Mandelbaum\altaffilmark{5}}

\altaffiltext{1}{Universit\"at Heidelberg, Zentrum f\"ur Astronomie, Institut f\"ur Theoretische Astrophysik, Philosophenweg 12, 69120 Heidelberg, Germany; adizitrin@gmail.com}
\altaffiltext{2}{Rutgers University, Department of Physics \& Astronomy, 136 Frelinghuysen Rd, Piscataway, NJ 08854, USA}
\altaffiltext{3}{Space Telescope Science Institute, 3700 San Martin Dr, Baltimore, MD 21218, USA}
\altaffiltext{4}{Departamento de Astronom\'ia y Astrof\'isica, Facultad de F\'isica, Pontificia Universidad Cat\'olica, Casilla 306, Santiago 22, Chile}
\altaffiltext{5}{Department of Physics, Carnegie Mellon University, Pittsburgh, PA 15213, USA}
\altaffiltext{*}{\textbf{The mass model and parameters are publicly available at: ftp://wise-ftp.tau.ac.il/pub/adiz/ElGordo}}
\begin{abstract}
We present the first strong-lensing (SL) analysis of the galaxy cluster ACT-CL J0102-4915 (\emph{El Gordo}), in recent \emph{HST}/ACS images, revealing a prominent strong lens at a redshift of $z=0.87$. This finding adds to the already-established unique properties of \emph{El Gordo}: it is the most massive, hot, X-ray luminous, and bright Sunyaev-Zeldovich effect cluster at $z\gtrsim0.6$, and the only `bullet'-like merging cluster known at these redshifts. The lens consists of two merging massive clumps, where for a source redshift of $z_{s}\sim2$ each clump exhibits only a small, separate critical area, with a total area of $0.69\pm0.11\sq\arcmin$ over the two clumps. For a higher source redshift, $z_{s}\sim4$, the critical curves of the two clumps merge together into one bigger and very elongated lens (axis ratio $\simeq5.5$), enclosing an effective area of $1.44\pm0.22\sq\arcmin$. The critical curves continue expanding with increasing redshift so that for high-redshift sources ($z_{s}\gtrsim9$) they enclose an area of $\sim1.91\pm0.30\sq\arcmin$ (effective $\theta_{e}\simeq46.8\pm3.7\arcsec$) and a mass of $6.09\pm1.04\times10^{14}M_{\odot}$. According to our model, the area of high magnification ($\mu>10$) for such high redshift sources is $\simeq1.2\sq\arcmin$, and the area with $\mu>5$ is $\simeq2.3\sq\arcmin$, making \emph{El Gordo} a compelling target for studying the high-redshift Universe. We obtain a strong lower limit on the total mass of \emph{El Gordo}, $\gtrsim1.7\times10^{15}M_{\odot}$ from the SL regime alone, suggesting a total mass of, roughly, $M_{200}\sim2.3\times10^{15}M_{\odot}$. Our results should be revisited when additional spectroscopic and \emph{HST} imaging data are available.

\end{abstract}

\keywords{dark matter, galaxies: clusters: individuals: ACT-CL J0102-4915, galaxies: clusters: general, galaxies: high-redshift,
  gravitational lensing: strong}


\section{Introduction}\label{intro}

Massive galaxy clusters at high redshifts are rare beasts that hold
important clues to the evolution of structure in the Universe, and can help probe the
current $\Lambda$CDM paradigm \citep[e.g.][]{HarrisonHotchkiss2012,Waizmann2012JeanClaude0717,Zitrin2009_macs0717}.

In this study we focus on ACT-CL J0102-4915, \emph{El Gordo}, a high redshift ($z=0.87$), massive cluster discovered by the Atacama Cosmology Telescope (ACT) as the most significant Sunyaev-Zeldovich \citep[SZ;][]{SunyaevZeldovich1972} decrement in their survey area of $\sim$1000 deg$^{2}$ \citep{Marriage2011ACT,Menanteau2012Gordo,Hasselfield2013ACT}. The cluster was also detected by the South Pole Telescope (SPT) in their 2,500 deg$^{2}$ survey as the highest significance SZ detection \citep{Williamson2011SPT}. Additionally, recent results by the \citet[][see Fig. 29 and related catalog therein]{2013PlanckSZsources} confirm that \emph{El Gordo} is an extreme case, with the highest SZ-estimated mass at $z\gtrsim0.65$. \citet{Menanteau2012Gordo} pursued an efficient multi-wavelength follow-up using the Very Large Telescope (VLT), \emph{Chandra} and \emph{Spitzer}. The spectra of 89 member galaxies yielded a cluster redshift, $z = 0.870$, and a velocity dispersion,
$\sigma_{gal} = 1321\pm106$ km s$^{-1}$. Their \emph{Chandra} observations revealed a hot ($kT_{X} =14.5\pm 1.0$ keV) and X-ray luminous ($L_{X} = 2.19 \pm 0.11 \times 10^{45}$ erg s$^{-1}$) system with a complex morphology \citep[see][]{Menanteau2012Gordo}; these values place \emph{El Gordo} at the extreme massive end of all known clusters.

\citet{Menanteau2012Gordo} determined the mass of \emph{El Gordo} to be $M_{200} = 2.16 \pm 0.32 \times 10^{15} ~ M_{\odot}$, using multiple proxies such as the SZ effect, X-ray, and dynamics, making it the most massive and X-ray luminous galaxy cluster known at $z > 0.6$. Additionally, \emph{Chandra} and VLT/FORS2 optical data revealed that \emph{El Gordo} is undergoing a major merger between two components with a mass
ratio of approximately 2 to 1; the most plausible direction for the merger inferred from the
structures seen in the X-ray emission \citep[see][]{Menanteau2012Gordo} is along the NW-SE axis. To our knowledge, \emph{El Gordo} is the only ``Bullet-like" cluster known to date at $z>0.6$.

Due to the improvement of lens modeling techniques, Hubble Space Telescope (\emph{HST}) imaging and the studies of the high redshift Universe through cluster lenses \citep[e.g.][]{Kneib2004z7,Bradley2008,Bradley2011,Richard2011A383highz,Bradac2012highz,Zheng2012NaturZ,Zitrin2012CLASH0329,Coe2012highz}, the lensing efficiency and magnification power of galaxy clusters have been increasingly studied in recent years, in pursuit of the best cosmic telescopes \citep[e.g.][]{OguriBlandford2009,Fedeli2010,Meneghetti2010a,Redlich2012MergerRE,Wong2012OptLenses,Zitrin2012UniversalRE,Zitrin2013M0416}. The efficiency and magnifying capabilities of a lens depend on a variety of factors, such as the mass, ellipticity, concentration, mass profile, amount of substructure and its distance from the center, and degree of relaxation or merger, as well as the redshift of the lens (and source). Recent efforts have now established that there exists a particular class of prominent strong lenses, consisting of massive, mostly merging clusters (being found at increasing redshifts) for which the critical curves of the several mass clumps and different substructures can merge together into a bigger lens, resulting also in a shallower inner mass profile with higher magnification power. In addition, it has also been shown that high elongation of the critical curves boosts the lensing efficiency since the source-plane caustics get relatively bigger, generating more multiple image systems. We refer the reader to the following examples and related comprehensive studies  \citep[e.g][]{Fedeli2010,Meneghetti2010a,Lapi2012EllipLenses,Giocoli2012MOKA,SerenoZitrin2012,Paraficz2012Bullet,Redlich2012MergerRE,Wong2012OptLenses,Zitrin2009_macs0717,Zitrin2012UniversalRE,Zitrin2013M0416}.

For example, following these conclusions, such merging or substructured clusters have been now prioritized and chosen for the pioneering Frontier Fields\footnotemark[1] \footnotetext[1]{http://www.stsci.edu/hst/campaigns/frontier-fields/} program, set to detect the highest redshift galaxies magnified by cosmic telescopes, with the \emph{HST}.

Here, we present the first strong lensing (SL) analysis of \emph{El Gordo}, in recent \emph{HST} imaging with the Advanced Camera for Surveys (ACS). This cluster fulfills the criteria of the strongest lenses known experiencing a major merger between two massive clumps and has a very elongated mass distribution, yet has an additional intriguing property: its high redshift ($z_{l}=0.87$). For this lens redshift, the relative lensing distance, i.e. lens-to-source angular diameter distance over the source angular diameter distance, or $D_{ls}/D_{s}$, increases substantially for higher redshift sources (relative to lower redshift sources) resulting in, relatively, rapidly expanding critical curves, forming a useful lens for observing the high-$z$ Universe, as we shall show below. Throughout we adopt a concordance $\Lambda$CDM cosmology with ($\Omega_{\rm m0}=0.3$, $\Omega_{\Lambda 0}=0.7$, $H_{0}=100$ $h$ km s$^{-1}$Mpc$^{-1}$, with $h=0.7$), where $1\arcsec= 7.71$ kpc at the redshift of \emph{El Gordo}.

\begin{figure*}
 \begin{center}
   \includegraphics[width=150mm]{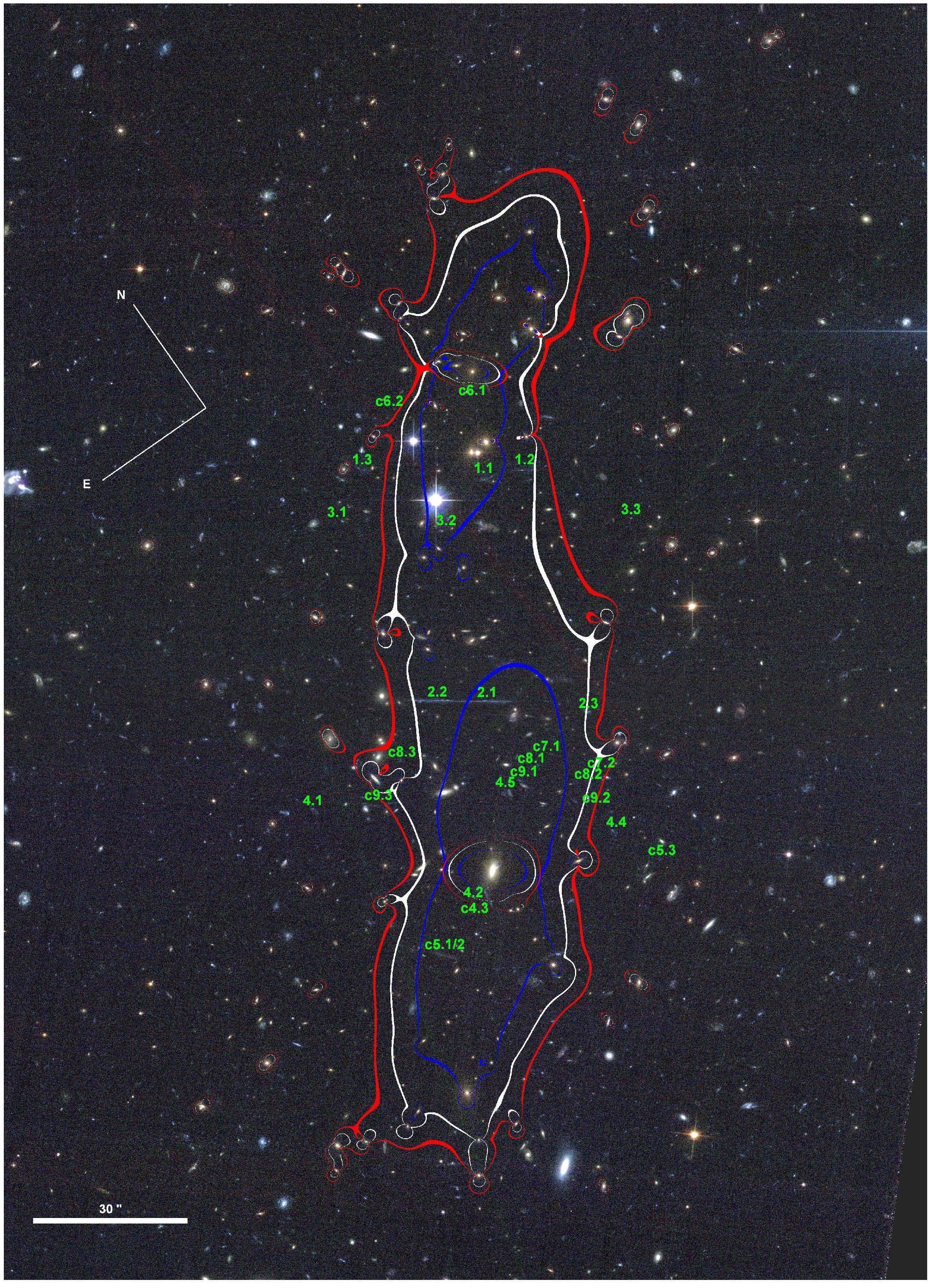}
 \end{center}
\caption{Critical curves for different source redshifts overlaid on a 3-band \emph{HST} image of \emph{El Gordo} ($RGB=[F850LP,F775W,F625W]$; North is 35 degrees counterclockwise of the positive $y$-axis). Using our unique LTM mass modeling technique \citep{Zitrin2009_cl0024} we have been able to physically find the first 27 multiple images and candidates of at least 9 background sources, as labeled on the image (``c" stands for candidate). We then used these multiple-image constraints to model the cluster as two $e$NFW halos representing the DM, plus PIEMD parameterizations for the galaxies. The resulting critical curves overlaid in \emph{blue} correspond roughly to systems 1, 2, and 4, at a typical redshift of $z_{s}\sim2-2.5$.  With increasing redshift the critical curves of the two clumps merge together into one bigger lens. Overlaid in \emph{white} are the critical curves for a source at $z_{s}\sim4$ (system 3), and the outer \emph{red} critical curve corresponds to a high-redshift, $z_{s}\sim9$ source.}
\label{multipleimages}
\end{figure*}

\begin{figure}
 \begin{center}
   \includegraphics[width=80mm]{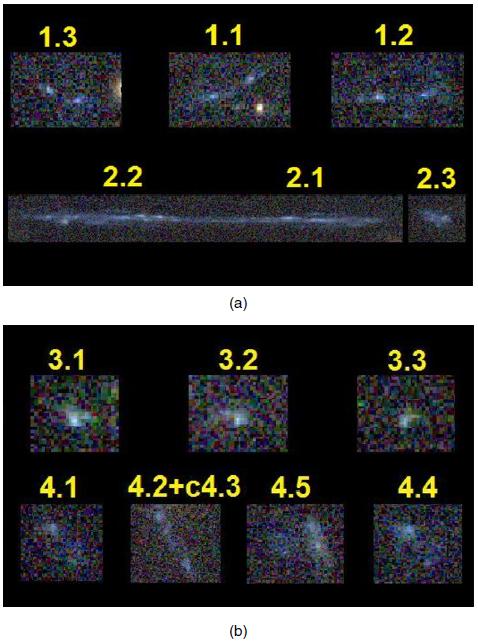}
 \end{center}
\caption{Zoomed-in examples of multiple images identified.}
\label{multipleimages2}
\end{figure}

\section{Observations and Data reduction}\label{Obs}
\emph{El Gordo} was imaged with the ACS on board the \emph{HST} in the F625W, F775W, and F850LP bands, on 2012 Sep 19 (prop ID: 12755, PI: Hughes), with integration times of 2344s, 2512s, and 2516s, respectively. The ACS dataset consists of two contiguous
pointings centered on the NW and SE clumps of the cluster with a rotation angle of approximately $55\deg$. The data were first processed by the STScI ACS Calibration pipeline (CALACS), which included bias and dark subtraction,
flat-fielding, counts-to-electrons conversion and charge transfer
efficiency (CTE) correction using the Pixel-based method described in
\cite{Ubeda2012}.

The images were then processed using an updated version of APSIS \citep{Blakeslee2003APSIS}, to reject cosmic rays (CRs) and combine the images in
each band into geometrically corrected, single-field drizzled images. Object detection, extraction, and integrated photometry, were performed using
SExtractor \citep{BertinArnouts1996Sextractor} catalogs, produced by the Apsis
pipeline. Photometric redshift estimates were computed with the
Bayesian Photometric Redshift package \citep[BPZ;][]{Benitez2000} using
isophotal-corrected magnitudes, and the prior calibrated with the
northern HDF (HDF-N) spectroscopic sample. 

We also make partial use here of the discovery observations of \emph{El Gordo}, which took place in 2009 December with the 4.1 m SOAR Telescope using the $griz$ filter set, with exposure times of 540s, 720s, 2200s, and 2200s, respectively, and a typical seeing of $<0''\!.7$ \citep[][for complete details]{Menanteau2012Gordo}. Subsequently, \emph{El Gordo} was observed using the FORS2 on VLT providing redshifts for 123 objects (89 cluster members). In addition to the \emph{HST} images which are our primary dataset for the lensing analysis, we use these other data both to help choose cluster members and visually inspect multiple image candidates, particularly in the SOAR $g$-band data, which cover bluer wavelengths than the ACS F625W filter.

\section{Strong Lensing Analysis of \emph{El Gordo}}\label{SLanalysis}

For the SL analysis of \emph{El Gordo}, we first use the method outlined in \citet{Zitrin2009_cl0024} which adopts a light-traces-mass (LTM) assumption for both the galaxies and dark matter (DM), where the latter is simply a smoothed version of the former, and the two components are added together and supplemented by an external shear to allow for more freedom and higher elongation \citep[see][for full details]{Broadhurst2005a,Zitrin2009_cl0024}. This method has been successfully applied to a large number of clusters \citep[e.g.][]{Zitrin2011_12macsclusters,Zitrin2012CLASH1206,Zitrin2013M0416,Merten2011,Coe2012highz,Zheng2012NaturZ}. Thanks to the LTM assumption and the low number of parameters, the initial model is already constrained well enough to aid in finding, physically, multiple images across the cluster field, which are then used to iteratively refine the model. With this method, along with a complementary examination by eye, we uncovered in \emph{El Gordo} four secure sets of multiple images and five additional multiply-imaged candidate systems.

After physically matching up multiple images with the LTM model, we then model the cluster with a more flexible parametrization, consisting of two elliptical NFW halos ($e$NFW) representing the two cluster-scale DM clumps, and adopting Pseudo-Isothermal Elliptical Mass Distributions (PIEMD) for the galaxies. This parametrization consists of a total of 10 fundamental parameters: $r_{cut}^{*}$ and $\sigma_{0}^{*}$, the cutoff radius and velocity dispersion of a typical $L^{*}$ galaxy, for the PIEMD galaxy models \citep[e.g.][]{Jullo2007Lenstool}; the scale radius $r_{s}$ and the concentration parameter $c_{vir}$, as well as the ellipticity and its position angle, for each of the two $e$NFW halos, whose centers are fixed on the central galaxies of the SE and NW clumps, respectively. The best fit solution is obtained via a long (several dozens of thousands steps) Monte-Carlo Markov Chain (MCMC) minimization. We previously used this method in our recent work, where more complete details can be found \citep[][and references therein]{Zitrin2013M0416}.

In total, we found 27 multiple images and candidates of 9 background sources (Table \ref{multTable}, Figs. \ref{multipleimages} and \ref{multipleimages2}), all of which were previously unknown, except for the giant arc (images 2.1+2.2) noted by \citet{Menanteau2012Gordo} as a possible multiply-lensed galaxy. All images \emph{not} marked as candidates were used as constraints for the model: 25 images and internal distinctive knots, of 4 sources. The multiple images uncovered are well detected in the bluest \emph{HST}, F625W band, and in the bluer, ground-based $g$-band (\S \ref{Obs}) and thus cannot be at redshifts larger than $z_{s}\sim4$. Due to the high redshift of the cluster, it is also not likely that they are at lower redshifts than $z_{s}\sim1.5$, since the lensing efficiency for lower source redshifts is very low. System 3 is only marginally detected in the $g$-band and thus is probably at a higher redshift than most of the other systems, around $z_{s}\sim4$, as also supported by our initial LTM model. Using this fact as a prior on the photo-$z$ of this system, we obtain a combined redshift and $95\%$ C.L. of 4.16 [4.04--4.23] for system 3. These constraints allow us to construct a SL mass model with Bayesian estimates for the model variables, including the redshifts of the different multiply-lensed sources. We thus fix the redshift of system 3 to $z_{phot}=4.16$, and allow the redshifts of the other systems (1, 2, and 4) to vary around their median photo-z value, with a flat prior. The resulting model redshift estimates are given in Table \ref{multTable}.

For the final model, the image-plane reproduction rms($\chi^{2}$) is 3.2\arcsec(122.66), using only images (and their distinctive knots) referred to as secure. For the $\chi^{2}$ we used a positional error of $\sigma_{pos}=1.4\arcsec$, which was found to be a reasonable value accounting for large-scale structure and matter along the line-of-sight \citep[][]{Jullo2010,D'AloisioNatarajan2011LOS,Host2012LOS}. The multiple-images input comprises 25 constraints, where the number of DOF is 23 yielding correspondingly $\chi^{2}/DOF$=5.3. These values are slightly higher than similar detailed lensing analyses, but are typical of complex or merging systems \citep[e.g.][]{Broadhurst2005a,Limousin2012_M0717}. Also, note that due to the relatively low number of constraints we did not leave the BCGs or other bright members to be freely weighted, nor allowed the NFW halo centers to vary, which often refine the fit. We note that with further careful analysis and additional \emph{HST} observations particularly with a wider color range, more multiple image systems are anticipated to be uncovered, and the mass model could be improved further.

\begin{figure}
 \begin{center}
   \includegraphics[width=90mm]{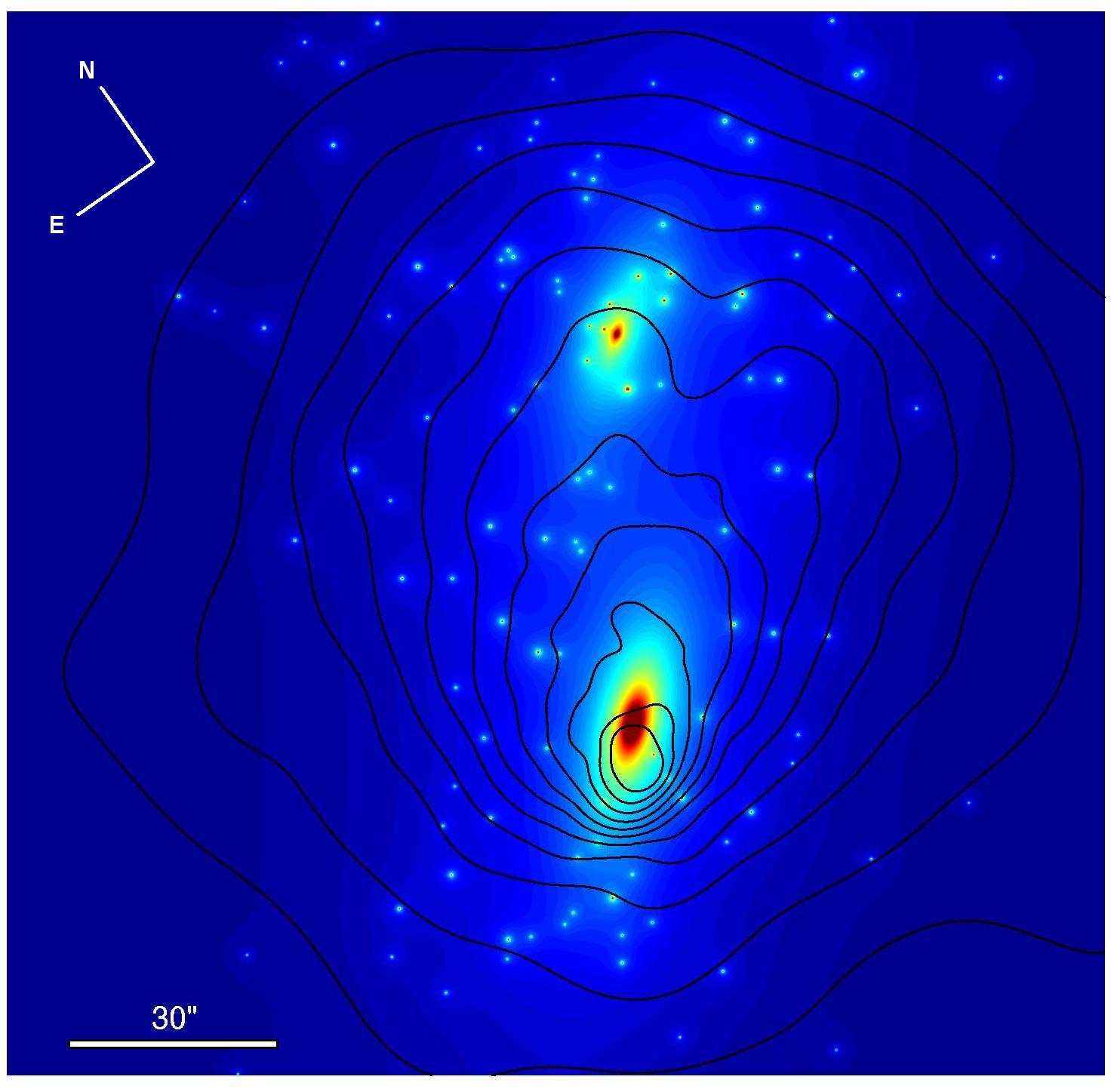}
 \end{center}
\caption{Surface density ($\kappa$) map of \emph{El Gordo} (arbitrary color scale). Orientation is identical to that of Fig. \ref{multipleimages}. Overlaid in \emph{black} are the \emph{Chandra} X-ray surface brightness contours from \citet[][see also for discussion on the observed offsets from the X-ray peak]{Menanteau2012Gordo}.} 
\label{kappamap}
\end{figure}

\begin{table*}\tiny
\caption{Multiple Images and Candidates found by our LTM method}\footnotesize
\label{multTable}
\centering
\begin{tabular}{|c|c|c|c|c|c|c|}
\hline
Arc ID & RA(J2000.0) & DEC(J2000.0)  & phot-$z$ [$z_{min}$-$z_{max}$] & $z_{model}$ & $\Delta$ pos $\arcsec$ & comments\\
\hline\hline
1.1a & 01:02:53.293 & -49:15:16.13 & 1.800 [1.613--2.810]  &  2.69 [1.15--3.38] & 1.1& \\
1.2a & 01:02:52.837 & -49:15:18.02 & 1.700 [1.520--2.980]  &  --- & 4.8 & \\
1.3a & 01:02:55.422 & -49:14:59.69 & 2.380 [1.880--2.690]  &  --- & 5.1 & \\
1.1b & 01:02:53.340 & -49:15:16.00 & 1.180 [1.035--2.400]  &  --- & 1.0 & \\
1.2b & 01:02:52.772 & -49:15:18.34 & 1.690 [1.511--2.320]  &  --- & 4.7 & \\
1.3b & 01:02:55.402 & -49:15:00.01 & 2.400 [1.880--2.710]  &  --- & 5.1 & \\
1.1c & 01:02:53.489 & -49:15:15.65 & 2.200 [1.987--3.310]  &  --- & 0.8 & \\
1.2c & 01:02:52.618 & -49:15:19.32 & 2.800 [2.470--3.053]  &  --- & 4.1 & \\
1.3c & 01:02:55.331 & -49:15:00.86 & 2.800 [2.540--3.053]  &  --- & 5.2& \\
\hline
2.1a & 01:02:55.861 & -49:15:51.94 & 2.210 [1.996--2.460]  &  2.11 [1.85--3.07] & 1.5& \\
2.2a & 01:02:56.760 & -49:15:45.58 & 3.140 [2.864--3.416]  &  " & 3.2& \\
2.3a & 01:02:54.418 & -49:16:04.20& 2.510 [2.276--2.750]  &  " & 2.4& \\
2.1b & 01:02:55.704 & -49:15:53.32 & --- &  " & 1.7 & \\
2.2b & 01:02:56.885 & -49:15:45.06 & --- &  " & 1.3 & \\
2.3b & 01:02:54.467 & -49:16:03.89 & --- &  " & 0.2 & \\
2.1c & 01:02:56.005 & -49:15:50.81 & --- &  " & 1.3 & \\
2.2c & 01:02:56.584 & -49:15:46.74 & --- &  " & 0.7 & \\
2.3c & 01:02:54.394 & -49:16:04.18 & --- &  " & 0.6 & \\
\hline
3.1 & 01:02:56.268 & -49:15:06.60 & 4.160 [3.816--4.504]   &  4.16 & 1.4 & sys fixed to $z_{phot}$=4.16 \\
3.2 & 01:02:54.760 & -49:15:19.18 & 1.100 [0.960--3.980]   &  " & 0.9& $z_{phot}$ 4.000 [3.420--4.333] using a prior (\S\ref{SLanalysis})\\
3.3 & 01:02:51.545 & -49:15:38.02&  0.900 [0.773--4.390]   &  " & 2.2& $z_{phot}$ 4.360 [4.003--4.717] using a prior (\S\ref{SLanalysis})\\
\hline
4.1 & 01:02:59.997 & -49:15:49.11 & 1.690 [1.511--3.310]  &  2.15 [1.86--2.91] & 6.1& \\
4.2 & 01:02:58.148 & -49:16:21.50 & 2.240 [2.024--3.040]  &  " & 3.0& \\
c4.3 & 01:02:58.178 & -49:16:24.10 & 2.100 [1.893--3.060]  &  " & 0.3& probable radial counter image\\
4.4 & 01:02:55.362 & -49:16:25.73 & 2.300 [2.080--3.120]  &  " & 1.9& \\
4.5 & 01:02:56.610 & -49:16:07.91 & 4.070 [3.732--4.408]  &  " & 3.6& blended with nearby galaxy\\
\hline
c5.1 & 01:02:59.612 & -49:16:26.16 & 0.940 [0.811--1.130]  &  2.21 [1.91--4.04] & 3.6& \\
c5.2 & 01:02:59.449 & -49:16:27.81 & 0.920 [0.792--1.330]  &  " & 1.7& \\
c5.3 & 01:02:54.942 & -49:16:35.58 & 1.200 [1.053--1.450]  &  " & 5.6& \\
\hline
c6.1 & 01:02:52.380 & -49:15:00.91 & 3.100 [1.720--3.373]  &  2.12 [1.65--3.66] & 4.7& \\
c6.2 & 01:02:54.167 & -49:14:54.54 & 0.950 [0.820--1.420]  &  " & 5.6& \\
\hline
c7.1 & 01:02:55.499 & -49:16:06.89 & 0.900 [0.773--1.120]  &  2.01 [1.32--3.75] & 2.6& \\
c7.2 & 01:02:54.938 & -49:16:14.42 & 0.900 [0.773--1.190]  &  " & 0.2& \\
\hline
c8.1 & 01:02:55.858 & -49:16:07.13 & 1.750 [1.567--2.760]  &  1.86 [1.58--2.59] & 2.8& \\
c8.2 & 01:02:55.222 & -49:16:15.78 & 1.900 [1.707--3.270]  &  " & 2.1& \\
c8.3 & 01:02:58.026 & -49:15:51.33 & 1.440 [1.190--2.270]  &  " & 0.8& \\
\hline
c9.1 & 01:02:56.309 & -49:16:07.53 & 2.190 [1.977--3.150]  &  1.95 [1.28--2.75] & 2.7& \\
c9.2? & 01:02:55.179 & -49:16:22.71 & 3.300 [2.320--3.587]   &  " & 2.2& \\
c9.2? & 01:02:55.652 & -49:16:17.11 & 2.220 [1.910--2.910]  &  " & 2.8& \\
c9.3 & 01:02:59.054 & -49:15:52.92 & 2.300 [1.960--2.810] &  " & 4.5& \\
\hline\hline
\end{tabular}\footnotesize
\tablecomments{$\emph{Column 1:}$ arc ID. ``c"-ID stands for \emph{candidate}, for which the model-predicted location, or identification \emph{by eye}, was ambiguous. For systems 1 and 2 we identify and use different parts of the images designated by their ID followed by a/b/c;  $\emph{Columns 2 \& 3:}$ RA and DEC in J2000.0; $\emph{Column 4:}$ photometric redshift and 95\% C.L.; $\emph{Column 5:}$ predicted and 95\% C.L. redshift by the model; $\emph{Column 6:}$ reproduction distance of the image from its observed location, using a mean source position; $\emph{Column 7:}$ comments.}
\end{table*}

\subsection{Results and Discussion}\label{comp}
In Fig. \ref{multipleimages} we show the resulting critical curves for different source redshifts. The critical area for systems 1, 2, and 4, estimated at a typical source redshift of $z_{s}\sim2-2.5$, consists of two separate critical curves encircling the NW and SE mass peaks. The mass enclosed within this total critical area of $0.69\pm0.11\sq\arcmin$, is $3.12\pm0.53\times10^{14}M_{\odot}$ (errors correspond to $1\sigma$). For system 3, at an estimated source redshift of $z_{s}\simeq4.16$, the critical curves of the two clumps merge together into one bigger lens, enclosing an area of $1.44\pm0.22\sq\arcmin$, and a mass of $5.06\pm0.86\times10^{14}M_{\odot}$. For a much higher source redshift, $z_{s}\sim9$ for example, the critical curves enclose an area of $1.91\pm0.30\sq\arcmin$ (effective Einstein radius of $\theta_{e}=\sqrt\frac{A}{\pi}\simeq46.8\pm3.7\arcsec$, or $\simeq361\pm29$ kpc), and a mass of $6.09\pm1.04\times10^{14}M_{\odot}$. Additionally, we find that for this source redshift, the area of high magnification ($\mu>10$) is $\simeq1.15\sq\arcmin$, and the area with $\mu>5$ is $\simeq2.25\sq\arcmin$. These numbers mark \emph{El Gordo} as a compelling cosmic lens for studying the high-redshift Universe.

Our mass model suggests a mass ratio of $\sim$1.5:1 between the SE and NW clumps, respectively (see Fig. \ref{kappamap}). Note that this is opposite to the mass ratio calculated in \citet{Menanteau2012Gordo}, which found by velocity dispersion measurements that the NW clump was approximately twice as massive as the SE clump. However, clearly, the overall properties of a halo cannot be deduced properly from the narrow SL regime alone, and so we leave further examination of the mass ratio to complementary weak-lensing studies.

To constrain the total mass of this cluster, we simply sum the mass within the FOV presented in Fig. \ref{kappamap}, obtaining a mass of $\sim1.7\times10^{15}M_{\odot}$ in that region which constitutes a strong lower limit for the total mass of this system. A simple NFW fit, for example, to the radial mass profile (of the mass map presented in Fig. \ref{kappamap}) centered on the optical midpoint between the two clumps, suggests an overall mass of $M_{200}\simeq2.3\times10^{15}M_{\odot}$. This simplified, rough estimate, is in good agreement with the multi-method estimates of \citet{Menanteau2012Gordo}; $M_{200} = 2.16 \pm 0.32 \times 10^{15} ~ M_{\odot}$.

Recent statistical studies \citep[e.g.][]{OguriBlandford2009,Fedeli2010,Meneghetti2010a,SerenoZitrin2012,Redlich2012MergerRE,Wong2012OptLenses,Zitrin2012UniversalRE}, and previous well-studied examples (e.g. MACS J0717.5+3745, \citealt{Zitrin2009_macs0717,Limousin2012_M0717}; the Bullet cluster \citealt{Bradac2006Bullet,Paraficz2012Bullet}; Abell 2744 \citealt{Merten2011}), have shown that massive merging clusters, can form efficient lenses due to the substructured, spread out mass distribution which boosts the critical area, but also entails, usually, an overall shallower mass profile enhancing the magnification power \citep[see also][]{Coe2012highz,Zheng2012NaturZ}. In addition, recently \citet{Zitrin2013M0416} showed that higher ellipticities enhance the lensing efficiency producing a larger number of multiple images per critical area, since the source-plane caustics are relatively bigger. Our analysis shows that \emph{El Gordo} has a highly elongated mass distribution, or critical curves, with an axis ratio of approximately $\simeq5.5$, implying that more multiple images are likely to be uncovered in its field with deeper space imaging. Although \citet{Zitrin2013M0416} have found a similarly high axis ratio for MACS J0416.1-2403, their numerical simulations also indicate that such high elongations are rare; only 4\% of clusters with $M_{vir}\geq6 \times10^{14} h^{-1}M_{\odot}$ exhibit such high elongations, highlighting the exceptional nature of \emph{El Gordo} as a cosmic lens.

Using an up-to-date luminosity function (see \S 8 in \citealt{Coe2012highz}) convolved with our lens model, we predict that \emph{El Gordo} should comprise a few dozen $z_{s}\sim4-6$ galaxies, roughly three $z_{s}\sim8$ galaxies, and possibly a few more at $z\gtrsim9$ (with observations as deep as e.g. 26.75 AB in \emph{HST}'s F160W band, $10\sigma$), taking into account two adjacent WFC3/IR pointings covering roughly the FOV seen in Fig. \ref{multipleimages}. This is a fair number of high-$z$ galaxies for a single cluster. For comparison, \citet{Bouwens2012highzInCLASH} found three $z_{s}\sim9-10$ galaxies over 19 \emph{Cluster Lensing And Supernova survey with Hubble} (CLASH; \citealt{PostmanCLASHoverview}) clusters with a similar depth of $\sim$27 AB, and about $\sim15$ $z_{s}\sim8$ galaxies are expected to be uncovered over the same fields (L. Bradley et al., \emph{in prep}).

\section{Summary}
We presented the first SL study of \emph{El Gordo}, in which we uncovered 27 multiple images and candidates of 9 background sources, revealing a prominent and highly elongated $z_{l}=0.87$ lens efficient for lensing high-redshift sources. The resulting critical curves expand relatively rapidly with source redshift as may be expected from the redshift of the cluster, reaching a critical area of about $\sim2\sq\arcmin$ for $z_{s}\gtrsim10$, and enclosing a mass of more than $6\times10^{14}M_{\odot}$. According to our model, for such high-$z$ sources the area of high magnification ($\mu>10$) is $\simeq1.2\sq\arcmin$, and the area with $\mu>5$ is $\simeq2.3\sq\arcmin$. This rare lens shows again, as recently appreciated by various works, that the lensing properties of merging clusters, are usually boosted (dependent on the mass, shape, and distance between the merging subclumps, e.g. \citealt{Zitrin2013M0416}; see also \citealt{Fedeli2010,Redlich2012MergerRE,Wong2012OptLenses}).

We obtained a strong lower mass limit for \emph{El Gordo} of $\sim1.7\times10^{15}M_{\odot}$ from the SL regime alone, suggesting, crudely, a total mass of $M_{200}\simeq2.3\times10^{15}M_{\odot}$ in agreement with \citet{Menanteau2012Gordo}, $M_{200} = 2.16 \pm 0.32 \times 10^{15} ~M_{\odot}$. The existence of such massive clusters, in particular at redshifts as high as \emph{El Gordo's} and above \citep[e.g.][]{Jee2009CL1226,Rosati2009,Gonzalez2012z1p75lens}, can probe and provide interesting insight on structure formation in $\Lambda$CDM \citep[e.g.][]{HarrisonHotchkiss2012,Waizmann2012JeanClaude0717}. Based on their aforementioned mass estimate, \citet{Menanteau2012Gordo} found that \emph{El Gordo}, by itself, did not pose a strong challenge to $\Lambda$CDM. However, they noted that a more accurate mass estimate would be required to test this conclusion; if the mass were 3$\sigma$ higher than their estimate, \emph{El Gordo} would no longer be predicted to exist in $\Lambda$CDM, over the whole sky, with a 95\% confidence level. In this relation, we note that future spectroscopy and space imaging of this cluster should help to uncover more sets of multiple images, obtain more accurate redshifts, refine the mass model, and enhance the reliability of our results. We leave further discussion on the rarity of \emph{El Gordo} to future papers including both strong and weak lensing, when the characteristics of this cluster over the entire observed spectrum and radius range, are well understood.

In addition to testing cosmological models, cluster physics and structure evolution, based on its lensing properties outlined in this work, \emph{El Gordo} is yet another compelling target to access the early Universe searching for the first galaxies, in particular in the era of recent $z\gtrsim10$ galaxy discoveries \citep[e.g.][]{Bouwens2011NaturZ10Gal,Bouwens2012highzInCLASH,Coe2012highz,Zheng2012NaturZ,Ellis2013Highz} and the upcoming \emph{HST} Frontier Fields science.

\section*{acknowledgments}
We greatly thank the reviewer of this work, Marceau Limousin, for valuable comments.
AZ is supported by contract research ``Internationale Spitzenforschung II/2-6'' of the Baden W\"urttemberg Stiftung. JPH and FM are supported by {\it Chandra} grant no.~GO2-13156X and {\it Hubble} grant no.~HST-GO-12755.01-A. LFB is supported by FONDECYT grant No. 1120676. ACS was developed under NASA contract NAS 5-32865. Results are based on observations made with the NASA/ESA Hubble Space Telescope, obtained from the data archive at
the Space Telescope Science Institute. STScI is operated by the
Association of Universities for Research in Astronomy, Inc. under NASA
contract NAS 5-26555. This work is in part based on observations made with ESO Telescopes at the Paranal Observatory under programme ID 086.A-0425.

\bibliographystyle{apj}
\bibliography{outDan2}

\end{document}